%% file: skeleton.tex
\documentclass{PoS}

\usepackage{subcaption}
\usepackage{amsmath}

\title{Jet properties and correlations in multi-jet topologies in
CMS}

\ShortTitle{Jet properties and multi-jet correlations}

\author{\speaker{A. Bermudez Martinez on behalf of the CMS Collaboration}\\
        Deutsches Elektronen-Synchrotron (DESY)\\
        E-mail: \email{armando.bermudez.martinez@desy.de}}

\abstract{We present measurements of multi-jet event properties, performed using proton-proton collisions data recorded by the CMS experiment. The jet charge and jet mass distributions are considered in addition to a measurement of the azimuthal angular correlations in 2- and 3-jet events. The measurements are compared to predictions including higher orders matched to parton shower and hadronization, together with predictions from semi-analytical calculations beyond next-to-leading logarithmic accuracy.}

\FullConference{XXVI International Workshop on Deep-Inelastic Scattering and Related Subjects (DIS2018)\\
		16-20 April 2018\\
		University of Kobe, Japan}

\begin{document}

\include{introduction}

\bibliography{skeleton}
\bibliographystyle{hieeetr}






\end{document}

%% file: introduction.tex

The accurate description of multi-jet events, which are produced at the CERN LHC, relies on a good description of multi-partonic radiation in perturbative quantum chromodynamics (QCD). This can be achieved by means of higher order calculations and parton shower (PS) simulations. In this report the measurements of the jet-charge and -mass distributions are presented. In addition the measurement of the angular correlation between the leading jets in 2- and 3-jet inclusive events are also discussed. The data were recorded by the CMS experiment~\cite{Chatrchyan:2008aa}.   


\textbf{\underline{Jet charge}}

One can assign a charge to a jet based on the charge of the particles clustered by the jet algorithm. The jet hadronic content should still carry information on the originating parton charge. The specific procedure in which a jet charge is built up from the individual charges of the particles in the jet is called "jet estimator". By using different estimators for the jet charge one might be able to discriminate between quark- and gluon-like jets. The jet charge is sensitive to the radiation contributing to the jet therefore it is sensitive to the PS models and the underlying event activity. The jet charge estimator is based on the momentum-weighted sum of the electric charge of the jet constituents. The measurement of the leading jet charge is performed in ranges of its transverse momentum $\rm{p}_T$ using dijet events. This analysis~\cite{Sirunyan:2017tyr} is carried out with data collected by the CMS experiment in proton-proton collisions at $\sqrt{s}=8$ TeV corresponding to an integrated luminosity of 19.7 fb$^{-1}$. Two definitions of jet charge are shown here:

\begin{align}
Q^\kappa = \frac{1}{(\rm{p}_T^{jet})^\kappa} \sum_{i} Q_i (\rm{p}_T^i)^\kappa  && Q_{L}^\kappa  =  \left.\sum_{i} Q_{i} \left(p^{i}_{\|}\right)^{\kappa} \middle/ \sum_{i} \left(p^{i}_{\|}\right)^{\kappa} \right. ,
\end{align}

The sums above are over the particles $i$ in the jet with $\rm{p}_T > 1$GeV.
The variable $\rm{p}_T^{jet}$ is the transverse momentum of the jet, $Q_i$ is the charge of the particle, and $\rm{p}_T^i$ is the magnitude of the transverse momentum of the particle relative to the beam axis. In the $Q_{L}^\kappa$ definition, the notation $p^{i}_{\|}  = \vec{p}^{i}\cdot \vec{p}_{jet} / | \vec{p}_{jet}|$ refers to the component of the transverse momentum of particle $i$ along the jet axis. The $\kappa$ parameter in the exponent of the particle momenta controls the relative weight given to low and high momentum particles contributing to the jet charge.

\begin{figure}
\begin{subfigure}{.32\textwidth}
  \centering
  \includegraphics[width=1.02\textwidth]{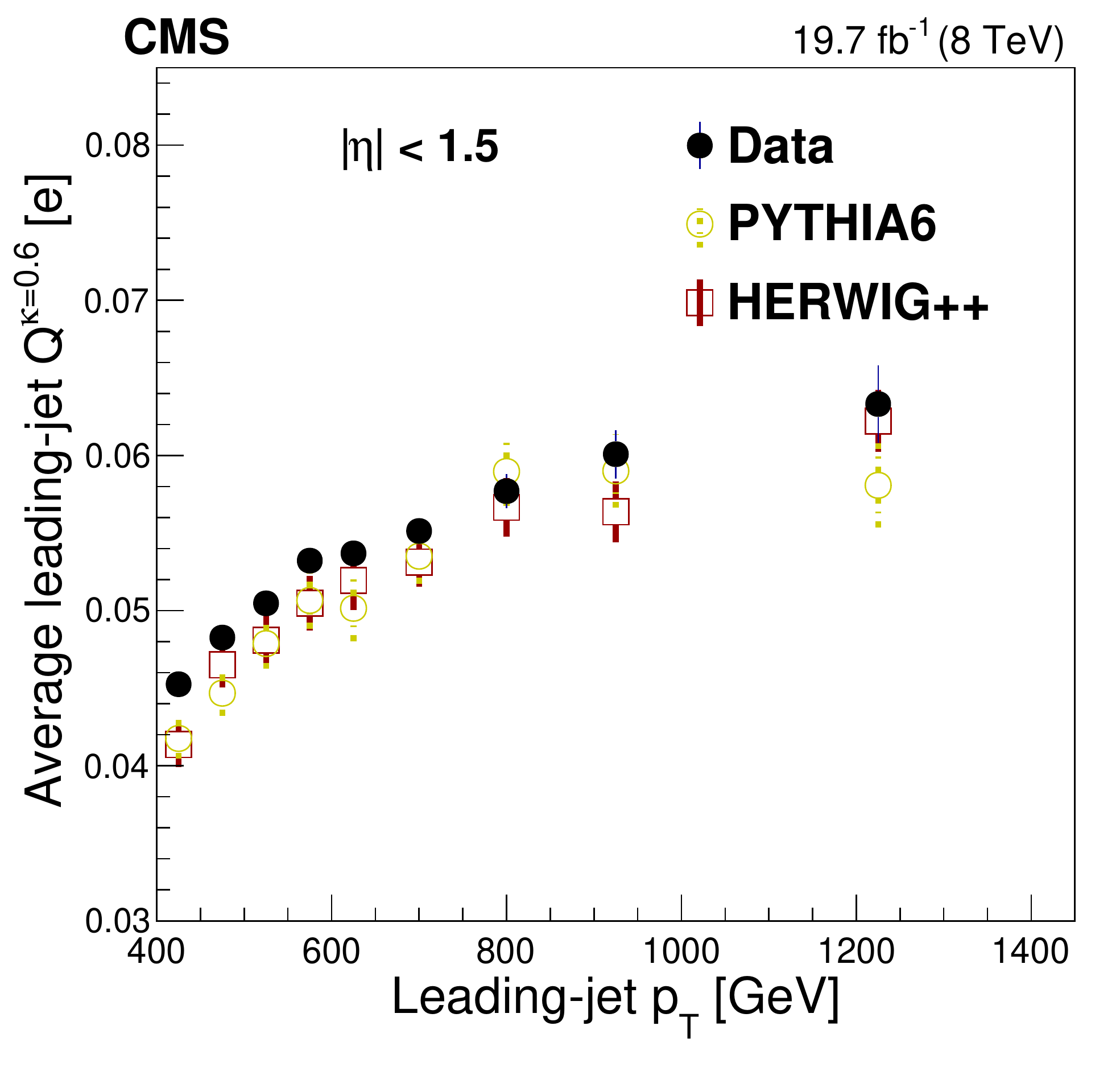}  
  \caption{1a}
  \label{fig:sfig1_1}
\end{subfigure}%
\begin{subfigure}{.32\textwidth}
  \centering
  \includegraphics[width=1.02\textwidth]{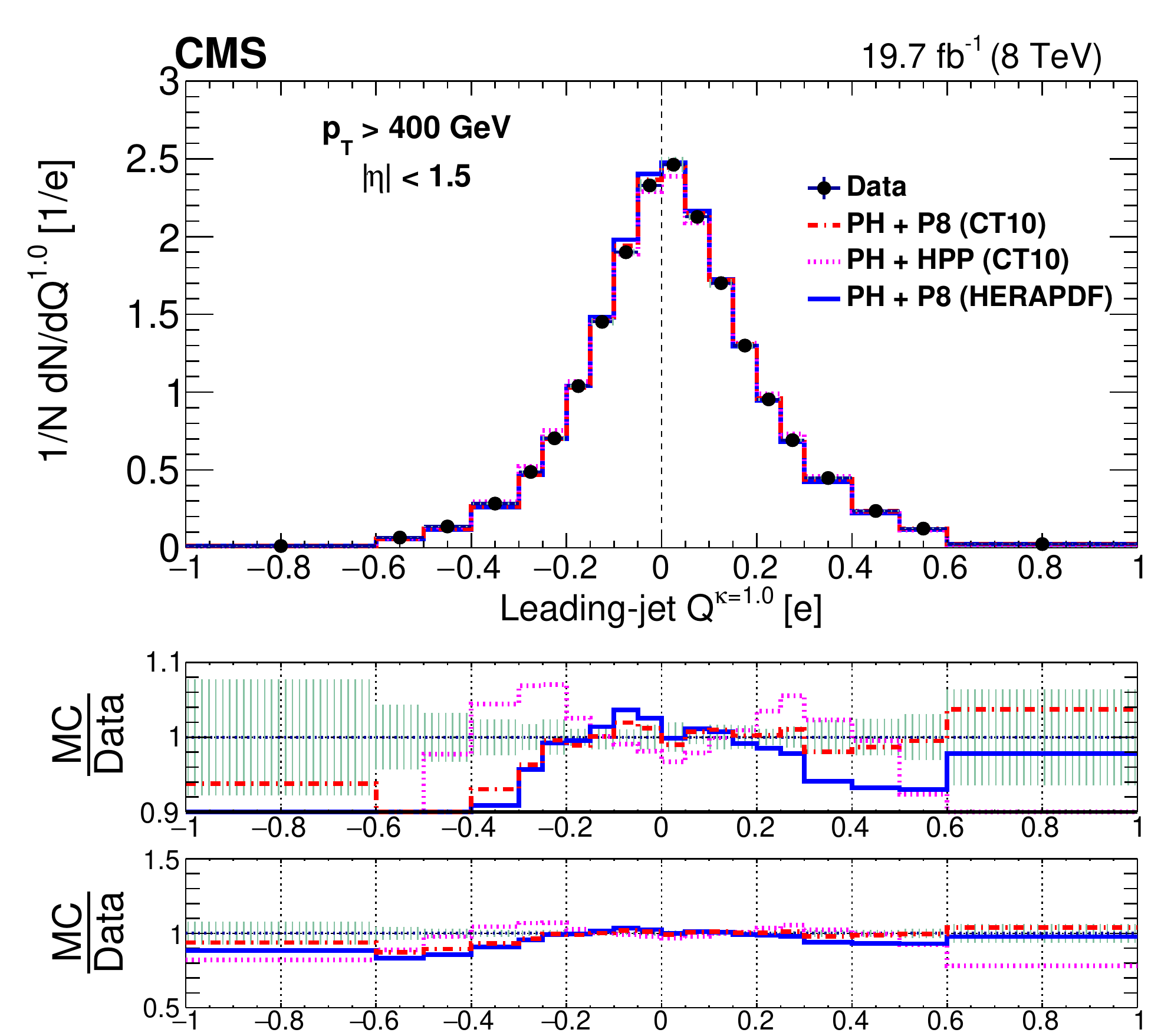}
  \caption{1b}
  \label{fig:sfig1_2}
\end{subfigure}
\begin{subfigure}{.32\textwidth}
  \centering
  \includegraphics[width=1.02\textwidth]{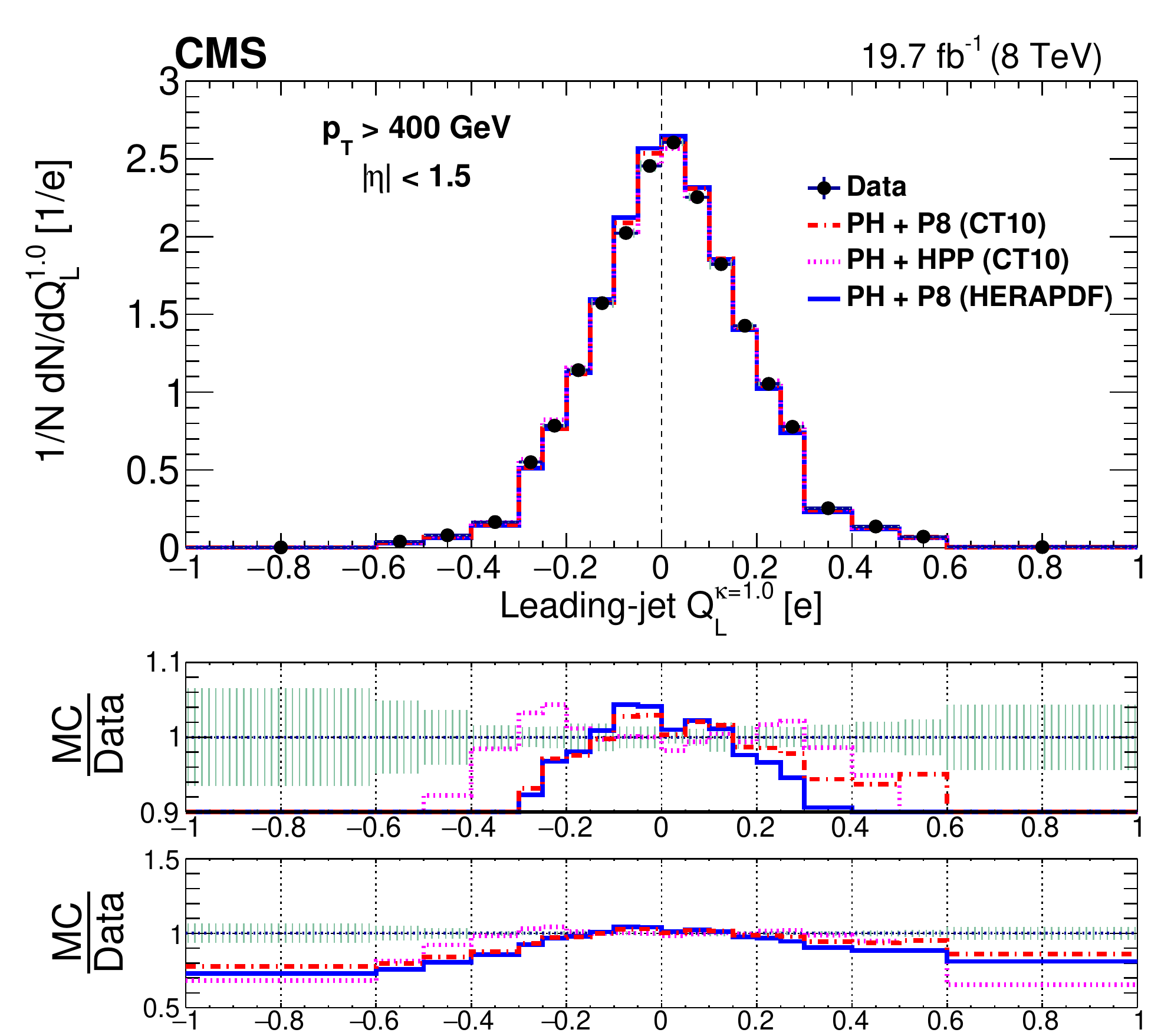}  
  \caption{1c}
  \label{fig:sfig1_3}
\end{subfigure}%
\caption{a) Data dependence of the average leading-jet charge $Q^{\kappa}$ with $\kappa = 0.6$ on the $\rm{p}_T$ of the leading jet and a comparison with simulations based on PYTHIA 6 and HERWIG++. Leading-jet charge distributions of $Q^{\kappa}$ (b) and $Q_L^{\kappa}$ (c) with POWHEG + PYTHIA8 (NLO CT10 and NLO HERAPDF 1.5 sets) and POWHEG + HERWIG++ generators are also shown.}
\label{fig:fig1}
\end{figure}

Gluon jets dominate the lower part of the $\rm{p}_T$ spectrum, while quark jets become progressively more relevant at high $\rm{p}_T$. As a consequence, and considering that the absolute value of the up-like quarks charge is bigger than the down-like quark one, the average jet charge with $\kappa=0.6$ increases as a function of the leading-jet $\rm{p}_T$, as can be observed in Fig.~\ref{fig:sfig1_1}. PYTHIA6~\cite{Sjostrand:2006za} (tune Z2$^{∗}$ ~\cite{Chatrchyan:2013gfi,Khachatryan:2015pea}) and HERWIG++~\cite{Bahr:2008pv} (tune EE3C ~\cite{Gieseke:2003rz}) simulations reproduce this trend. The dominant uncertainties arise from the track $\rm{p}_T$ resolution and the modelling of the response matrix.
The remaining systematic uncertainties have small effects (less than a percent) and include the jet energy scale and jet energy resolution. Figures~\ref{fig:sfig1_2} and ~\ref{fig:sfig1_3} present the distributions of the unfolded data compared to the generator-level POWHEG~\cite{Frixione:2007vw,Alioli:2010xd,Nason:2004rx} + PYTHIA8~\cite{Sjostrand:2007gs} (tune CUETP8M1~\cite{Khachatryan:2015pea}) and POWHEG + HERWIG++ predictions using the CT10~\cite{Lai:2010vv} and HERAPDF~1.5~\cite{Aaron:2009aa} NLO parton density sets with POWHEG + PYTHIA8.
The effect of the PS and fragmentation model on the jet charge distribution can be seen by comparing the predictions from POWHEG + PYTHIA8 with POWHEG + HERWIG++ simulations, which make predictions based on different models for PS and fragmentation. The effect of the parton density set on the jet charge distribution can be seen by comparing predictions with CT10 and HERAPDF~1.5. Simulations of initial-state radiation and multiple-parton interactions do not change the jet charge distribution. Disabling the simulation of final-state radiation in PYTHIA8, however, leads to a significantly broader jet charge distribution, which might mean that the jet charge distribution is mainly sensitive to the modelling of this effect. 

The use of jet charge estimators can help to discriminate between quark- and gluon-like jets. In addition the jet charge is sensitive to the radiation contributing to the jet therefore it can help improving the modelling of the PS and the underlying event activity.

\textbf{\underline{Jet mass}}

The mass of a jet can be calculated from the four-momenta of the particles entering the jet algorithm. Therefore the jet mass is not only sensitive to radiation from the originating parton but also to the underlying event activity, soft QCD contributions and radiation from additional pp interactions in the same bunch crossing (pileup (PU)). In this report a measurement of the differential jet production cross section as a function of the jet mass in bins of $\rm{p}_T$
from dijet events is presented. The analysis strategy presented is similar to that of Ref.~\cite{SMP-12-019}, except for the different center-of-mass energy of the collisions, and that both the jet $\rm{p}_T$ and $m$ are here corrected for detector effects. In order to distinguish the hard part contributing to the jet mass a jet grooming algorithm (soft drop (SD) mass~\cite{Larkoski:2014wba}) is applied that selectively removes low-energy portions from a jet. Comparing the production cross section with respect to the groomed and ungroomed jet mass separately allows us to
gain insight into both the hard and soft physics. The data have been recorded by the CMS Collaboration in proton-proton collisions at the LHC at a center-of-mass energy of $13$ TeV and correspond to an integrated luminosity of 2.3 fb$^{-1}$. The complete analysis can be found in~\cite{CMS:2017tdn}.

\begin{figure}
\begin{subfigure}{.5\textwidth}
  \centering
  \includegraphics[width=.8\textwidth]{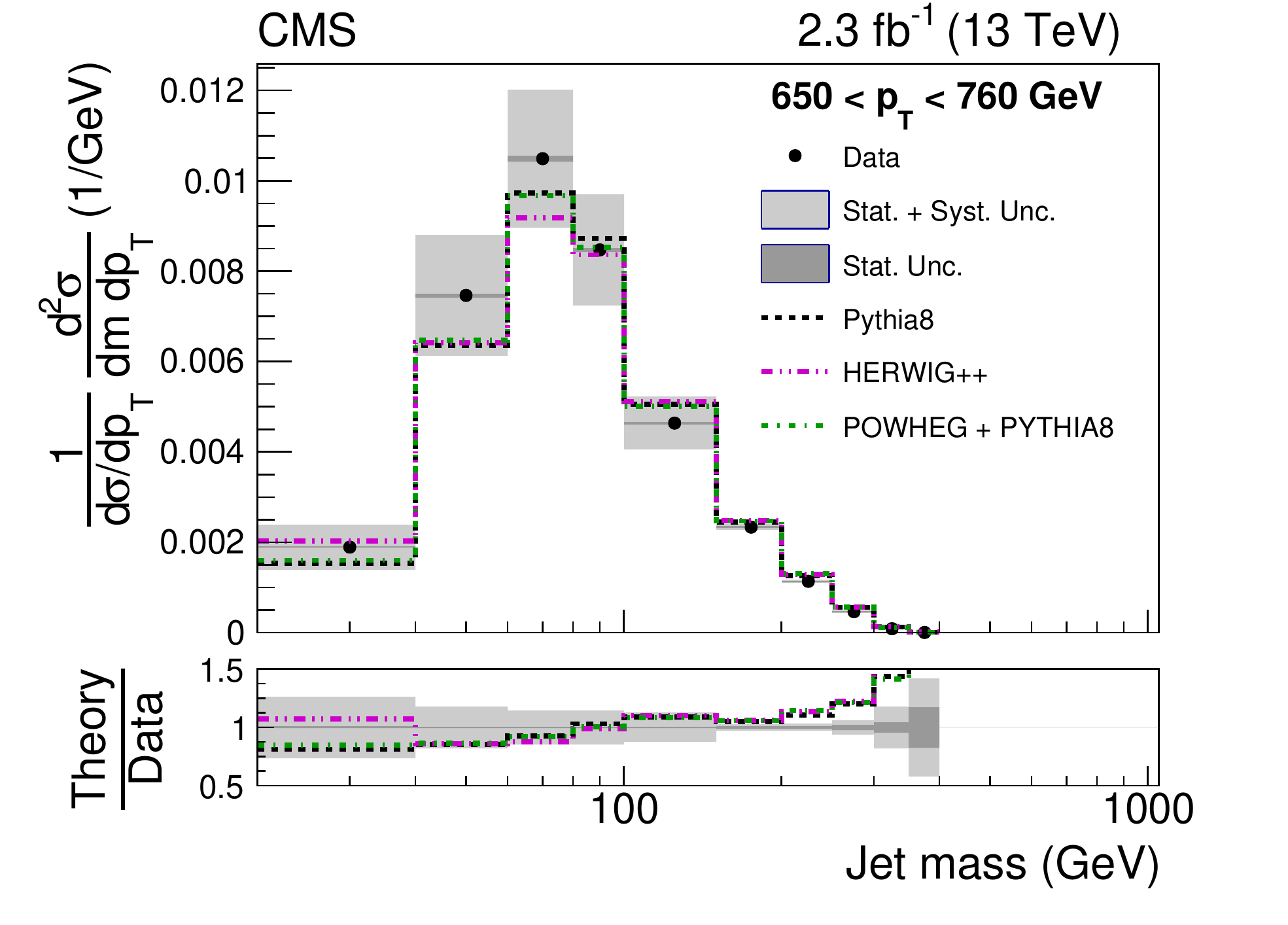}  
  \caption{1a}
  \label{fig:sfig2_1}
\end{subfigure}%
\begin{subfigure}{.5\textwidth}
  \centering
  \includegraphics[width=.8\textwidth]{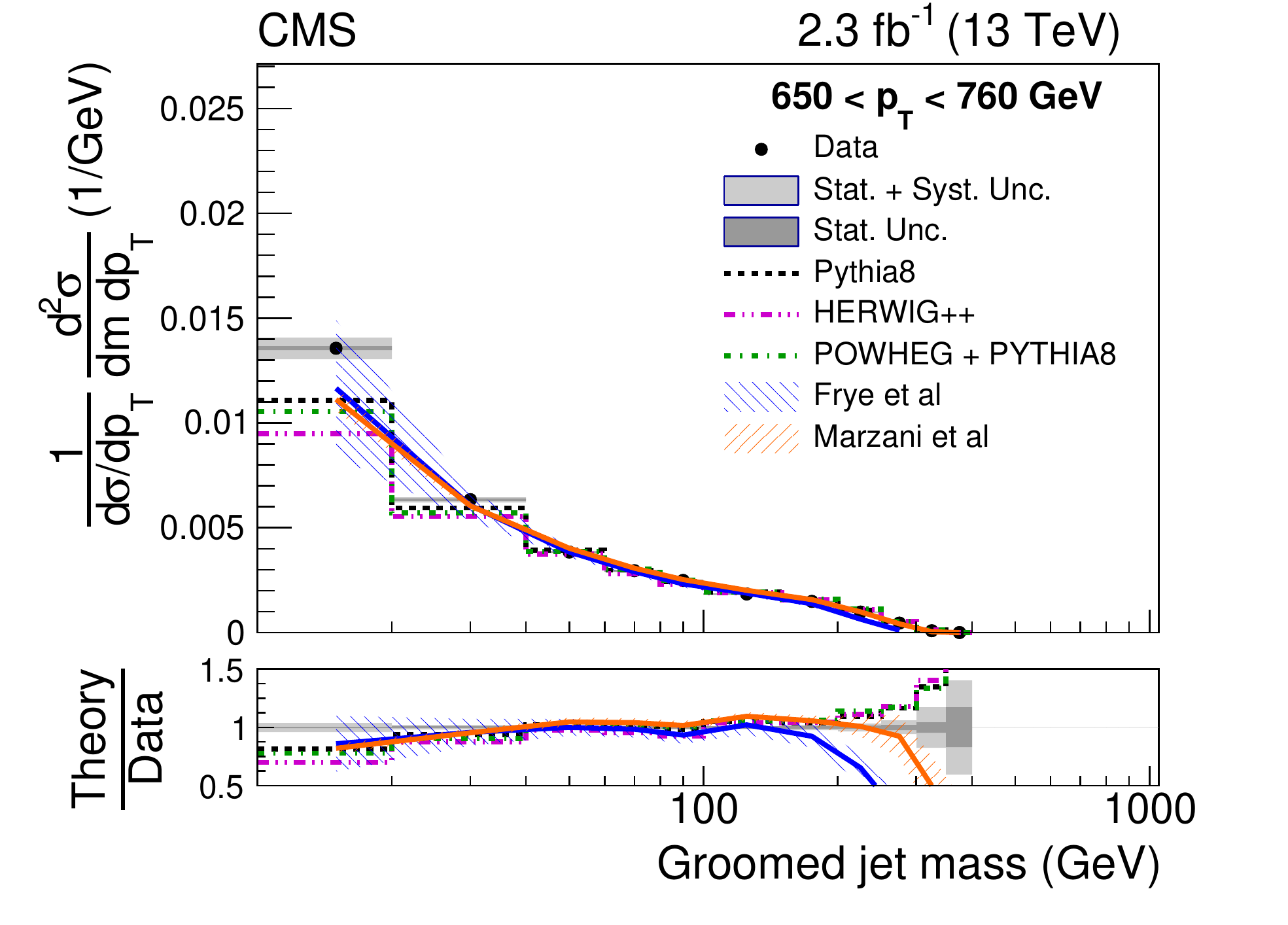}
  \caption{1b}
  \label{fig:sfig2_2}
\end{subfigure}
\caption{a) The predictions
from PYTHIA8, HERWIG++, and POWHEG+PYTHIA of the ungroomed jet mass compared to data. b) The predictions from PYTHIA8, HERWIG++, POWHEG+PYTHIA and the semi analytical predictions from Refs.~\cite{Frye:2016aiz} (scale variations and nonperturbative uncertainties included) and~\cite{Marzani:2017mva} (scale variations uncertainties included) are shown. }
\label{fig:fig2}
\end{figure}

In Fig.~\ref{fig:fig2} a measurement of the jet mass is shown. 
On the one hand Fig.~\ref{fig:sfig2_1} depicts the ungroomed jet mass and the predictions from PYTHIA8, POWHEG + PYTHIA8 (tune CUETP8M1), and HERWIG++ (tune CUETHS1~\cite{Khachatryan:2015pea}) event generators. On the other hand the groomed jet mass is shown in Fig.~\ref{fig:sfig2_2} and the comparison to the semi-analytical calculations from Refs.~\cite{Frye:2016aiz} and~\cite{Marzani:2017mva}. The SD algorithm considerably lowers the jet mass overall, as expected. The soft part of the jet is removed, leaving the hard part of the jet, which tends to have low mass. For ungroomed jets, Monte Carlo event generators are found to predict the jet mass spectrum within 20\% for masses in the range $0.1 < m/\rm{p}_T < 0.3$. Below the Sudakov peak, variations above 20\% are observed between the predictions and the data. Close to the jet splitting threshold, the predictions from PYTHIA8 and HERWIG++ agree with each other, but overshoot the data by 20-50\%. There is no significant difference observed when POWHEG + PYTHIA8 is used compared to PYTHIA8 alone. For groomed jets, the Sudakov peak is suppressed and the precision in the intermediate mass region $0.1 < m/\rm{p}_T < 0.3$ improves, since the grooming algorithm removes the soft portions of the jet arising from soft radiation. The PYTHIA8 generator tends to agree with the data better, although for $m/\rm{p}_T < 0.1$ , the agreement is worse than $0.1 < m/\rm{p}_T < 0.3$. The LO and NLO theory predictions with analytic resummation (Refs.~\cite{Frye:2016aiz} and~\cite{Marzani:2017mva}, respectively) agree overall with the data. The uncertainties at low jet mass are dominated by the physics model, jet mass scale, jet mass resolution, and PU uncertainties for the ungroomed jets. The grooming procedure reduces the physics model uncertainties by removing the soft components, as well as the PU uncertainty. At higher jet mass, the jets are at the threshold of splitting into two, so the uncertainties increase because of $\rm{p}_T$ bin migration. 

The jet mass observable is sensitive to the physics behind the modelling of the PS and underlying therefore it could be used in the tuning of the model parameters. The groomed mass measurement can serve as a test ground for comparing analytical calculations to those coming from PS event generators.

\textbf{\underline{Azimuthal correlations}}

Parton radiation also plays a major role in describing properly the azimuthal angular separation between the two highest $\rm{p}_T$ jets in the transverse plane, $\Delta\phi_{1,2}=\vert \phi_\mathrm{jet1}-\phi_\mathrm{jet2} \vert$. For $\Delta\phi_{1,2} \sim \pi$ a well defined description should account for the resummation of multiple soft gluon radiation. This is achieved by the PS description. In this report the measurement of the normalized inclusive 2- and 3-jet cross sections as a function of the azimuthal angular separation between the two leading $\rm{p}_T$ jets is presented. The measurement is done in several intervals of the leading jet $\rm{p}_T$ ($\rm{p}_T^{max}$ ), within the rapidity range $\vert y \vert<2.5$. This measurement is an extension of the work done in Ref.~\cite{Sirunyan:2017jnl}. A finer binning ($1^{\circ}$) was chosen and in order to study in detail the region $\Delta\phi_{1,2} \sim \pi$. Experimental and theoretical uncertainties are reduced by normalizing the $\Delta\phi_{1,2}$ distribution to the total dijet cross section. The inclusive 3-jet distributions differential in $\Delta\phi_{1,2}$ and $\rm{p}_T^{max}$, with the $\rm{p}_T$ of third highest $\rm{p}_T$ jet to be 1-2 orders of magnitude smaller than $\rm{p}_T^{max}$, is also suitable to test resummation effects arising from the presence of multiple scales in the interaction. The measurement is performed using data collected during 2016 with the CMS experiment at the CERN LHC, 
corresponding to an integrated luminosity of 35.9fb $^{-1}$ of proton-proton collisions at $\sqrt{s}=13$ TeV. 

\begin{figure}
  \centering
  \includegraphics[width=.65\textwidth]{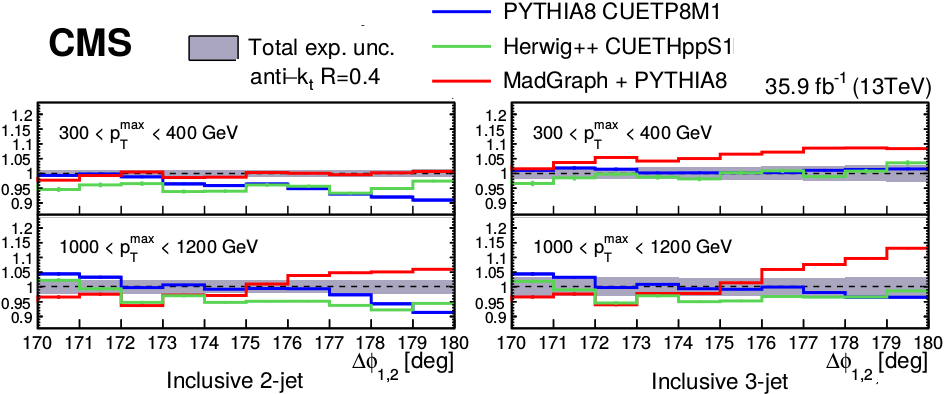}
\caption{Ratios of PYTHIA8, HERWIG++, and MADGRAPH + PYTHIA8 predictions to data, of the normalized inclusive 2- (left) and 3-jet (jet) distributions as a function of $\Delta\phi_{1,2}$, for two $\rm{p}_T^{max}$ regions. The experimental uncertainty (solid band) and the statistical uncertainty of the simulated data (error bars) are shown.}
\label{fig:fig3}
\end{figure}

Figure~\ref{fig:fig3} depicts the ratios of the predictions from PYTHIA8(tune CUETP8M1), HERWIG++(tune CUETHS1), and MADGRAPH~\cite{Alwall:2007fs,Alwall:2011uj} + PYTHIA8 
of the inclusive 2- (left) and 3-jet (right) jet distributions as a function of $\Delta\phi_{1,2}$ to data, for the different $\rm{p}_T^{max}$ regions. For the 2-jet case the event generators PYTHIA8 and HERWIG++ show the largest deviations from the measurements for the $\rm{p}_T^{max} < 800$ GeV regions in the inclusive 2-jet case, and the MADGRAPH + PYTHIA8 event generator gives the best description in the same regions. The three generators show large deviations from the measurements in the $\rm{p}_T^{max} > 800$ GeV regions. Opposite to the 2-jet case, for the 3-jet observables (right) MADGRAPH + PYTHIA8 shows the largest deviations from the measurements close to $180^{\circ}$ while PYTHIA8 and HERWIG++ give a good description of the data.
In general we observe that the $\Delta\phi_{1,2}$ region close to $180^\circ$ is not well described by the predictions. The fact that none of the generators is able to describe the 2- and 3-jet measurements simultaneously suggests that the observed differences (of the order of 10\%) are related to the the way soft partons are simulated within the PS. The observed differences between $\rm{p}_T$ and angular ordered PS for the LO generators PYTHIA8 and HERWIG++ are small.
The uncertainty from PS dominates for the normalized inclusive 2-jet distributions. 
On the other hand, for the normalized inclusive 3-jet distributions the main contributions come from PS and parton density uncertainties.

The measurement of angular correlations for collinear back$-­$to$-­$back dijet configurations probes the multiple scales involved in the event and therefore provides a sensitive test for models of soft parton radiation accompanying the hard process. Discrepancies between the measurement and theoretical predictions are as large as 15\% mainly in the region $177^\circ < \Delta\phi_{1,2} < 180^\circ$. The 2- and 3-jet measurements are not simultaneously described by any of models.

